\documentclass{article}

\usepackage{preprint}
\usepackage{upgreek}				
\usepackage[utf8]{inputenc} 		
\usepackage[T1]{fontenc}    		
\usepackage[dvipsnames]{xcolor}	
\usepackage{hyperref}       		
\hypersetup{
    colorlinks = true,
    linkcolor = {NavyBlue},		
    anchorcolor = .,
    citecolor = {NavyBlue},		
    filecolor = .,				
    menucolor = .,
    runcolor = {cyan}, 
    urlcolor = {NavyBlue},		
    }
\usepackage{url}            
\usepackage{booktabs}       
\usepackage{amsfonts}       
\usepackage{nicefrac}       
\usepackage{microtype}      
\usepackage{fancyhdr}       
\usepackage{graphicx}       
\usepackage{setspace} 
\usepackage{soul} 
\usepackage{ragged2e} 
\usepackage[toc,page]{appendix}
\usepackage{longtable}
\usepackage{eso-pic}
\usepackage{tikz}
\usepackage[superscript]{cite}

\graphicspath{{./Figures/}}     

  \centering
\title{A Compact Electrochemical Model for a Conducting Polymer Dendrite Impedance}

\author{
  Antoine Baron\textsuperscript{a}, Enrique Hern\'{a}ndez-Balaguera\textsuperscript{b}, S\'{e}bastien Pecqueur\textsuperscript{a}  \\
  \\
  a. IEMN, UMR 8520 \\
  Univ. Lille, CNRS, Univ. Polytechnique Hauts-de-France\\
  59000 Lille, France\\
  \\
  b. Universidad Rey Juan Carlos, M\'{o}stoles, Madrid\\ Matem\'{a}tica Aplicada, Ciencia e Ingeniería de los Materiales y Tecnolog\'{i}a Electr\'{o}nica\\
  \\
  \texttt{sebastien.pecqueur@iemn.fr} \\
}

\begin{document}
\maketitle

\begin{abstract}
Conducting Polymer Dendrites (CPD) are truly inspiring for unconventional electronics that shapes topological circuitries evolving upon an application. Driven by electrochemical processes, an electrochemical impedance rules signal propagation from one node to another. However, clear models dictating their behavior in an electroactive electrolyte have not been identified yet. In this study, we investigate on CPD in an aqueous electrolyte by impedance spectroscopy to unify their signal transport with an electrical model, aiming to define a circuit simulation block to integrate these objects in systems for \textit{in materio} information processing.

\end{abstract}

\raggedright

\justifying

\section{Introduction}

Conducting Polymer Dendrites (CPD) share interesting aspects of biology and electronics: conducting polymers are doped semiconductors enabling different technologies,\cite{Namsheer2021,Paulsen2020,CriadoGonzalez2021} and dendritic morphogenesis is reminiscent of metamorphic biological systems (roots, mycelia, brain cells) to exploit topological adaptation.\cite{Tero2010,Eichler2017,Simard1997} In digital electronics, information storage/processing is about polarizing electrodes to charge/discharge electrons in memories and open/close transistor switches. No mass transport is involved, so devices cannot heal, grow nor multiply. They can only be discarded when breaking, or when newer versions are released in the market. To enable such biological functionalities in electronics, one must investigate on mass-driven electrical processes for a computing paradigm where devices physically evolve to become better versions of themselves, via \textit{in operando} low energy/waste manufacturing thanks to on board electrochemistry. As such, CPD exploit electropolymerization to build electric interconnections,\cite{Dang2014,Koizumi2016,Ohira2017,Watanabe2018,Chen2023} performing synaptic and topological plasticity as adaptive mechanisms for classification.\cite{AkaiKasaya2020,Cucchi2021,Petrauskas2021,Janzakova2021,Hagiwara2023} Their growth mechanism is triggered by low voltage spike signals which oxidize a monomer into a conducting polymer on an electrode, whose morphology depends on the electrical information electrodes input to the monomer-loaded electrolyte.\cite{Eickenscheidt2019,Janzakova2021a} They have mostly been investigated in the time domain. Their electrochemical impedance behavior is still overlooked.\cite{Janzakova2021a,Scholaert2022}\\

Electrochemical impedance spectroscopy (EIS) is a well-established technique to understand charge transport mechanisms and interfacial phenomena in advanced functional materials such as conducting polymers. The impedance of polymer-based devices has previously been interpreted from the perspective of equivalent circuits involving classical circuit theory.\cite{Pecqueur2019} As a multi-level structure, fractality of CPD is expected to greatly condition the charge-discharge processes and degenerate furthermore the ionic charge migration, compared to a Debye relaxation ruling an ideal capacitor, which yields a distribution of time constants. The relationship between a CPD's morphology and its resulting electrochemical impedance has not been established yet, nor modeled in a unified circuit.\\

\section{Results and Discussion}

\subsection{Materials and methods}
\begin{figure}[h]
  \centering
  \includegraphics[width=0.8\columnwidth]{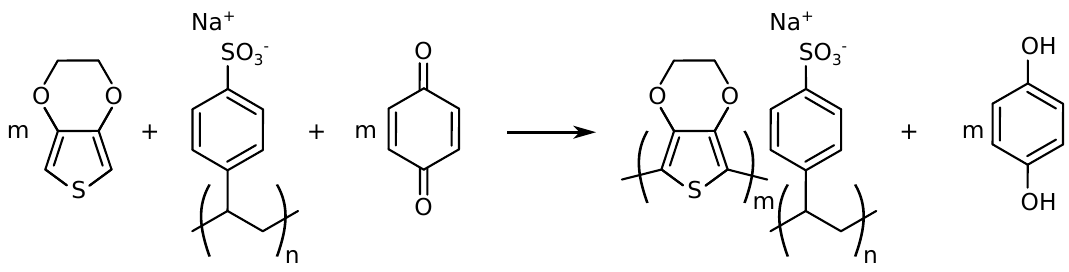}
  \caption{Electropolymerization of EDOT to PEDOT:PSS in water with NaPSS as counter-anion and BQ as oxidizing agent and proton scavenger.}
  \label{fig:fig1}
  \end{figure}

CPD have been grown by dynamic electropolymerization with a voltage source under pulse wave modulation. The electrolyte used was composed of 10~mM 3,4-ethylenedioxythiophene (EDOT, purchased from Sigma Aldrich), 10~mM para-benzoquinone (BQ, purchased from Sigma Aldrich), 1~mM of sodium polystyrene sulfonate (NaPSS, purchased from Sigma Aldrich) in deionized water. The experiment was performed in ambient, with two gold wires (purchased from Goodfellow) of 25~$\mu$m in diameter. At high frequency, EDOT electropolymerization in water is assumed to be dictated by the following reaction (Fig.~\ref{fig:fig1}):\\

\subsection{Conducting Polymer Dendrite Growths}
\begin{figure}[h]
  \centering
  \includegraphics[width=0.45\columnwidth]{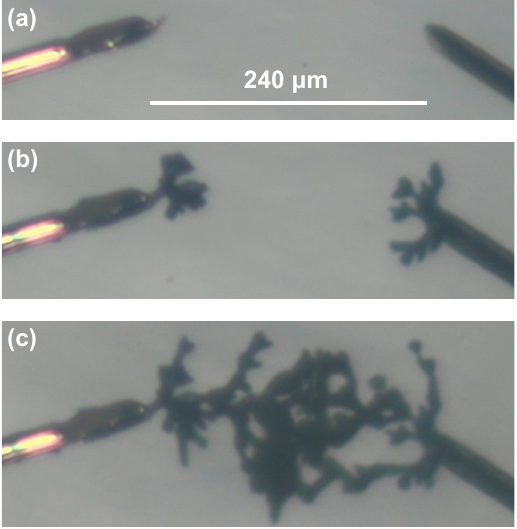}
  \caption{Microscope pictures of two adjacent gold wires immersed in an EDOT-containing electrolyte. The images were taken at the beginning of a growth (a), 2~min after start (b), and 6~min after start.}
  \label{fig:fig2}
  \end{figure}

CPD growths were performed with a voltage pulse-waveform. The characteristics of the stimuli were a peak voltage (\textit{V}$_{p}$) of 5~V, no voltage offset (\textit{V}$_{off}$), at 80~Hz frequency (\textit{f}), and 50\% of duty cycle (dc). A balance between structural disorder and growth directivity is systematically observed in all conducted experiments (Fig.~\ref{fig:fig2} as an example).\\

\subsection{Electrical Characterization and Modeling}
\begin{figure}[h]
  \centering
  \includegraphics[width=0.5\columnwidth]{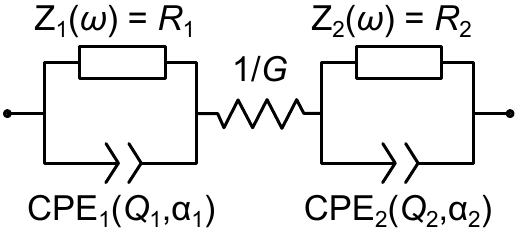}
  \caption{Equivalent circuit considered in this study.}
  \label{fig:fig3}
  \end{figure}

Frequency-resolved measurements were performed using a Solartron Analytical (Ametek) impedance analyzer (0~V offset, 10~mV$_{rms}$ amplitude) with a frequency ranging from 1~MHz to 100~mHz (71 discrete frequencies logarithmically spaced). The experimental impedance data were modeled using ZView and the following equivalent circuit (Fig.~\ref{fig:fig3}). The model consists of an electrolyte resistance 1/\textit{G} (in $\Upomega$) in series with two Voigt (R|CPE) subcircuits, each featuring an ohmic charge-transfer resistor (\textit{R}$_{i}$) in parallel with a constant-phase element (CPE$_{i}$). CPE$_{i}$'s impedance is given by \textit{Z}$_{CPE_{i}}$($\upomega$)~=~1/\textit{Q}$_{i}$(j$\upomega$)$^{\upalpha _i}$, where \textit{Q}$_{i}$ is a pseudo-capacitance, j$^{2}$~=~-1, $\upomega$ the angular frequency related to the frequency \textit{f} by $\upomega$~=~2$\uppi$\textit{f}, and $\upalpha_{i}$ (0~<~$\upalpha_{i}$~<~1) the dispersion coefficient. For each Voigt circuit, an effective capacitance \textit{C}$_{eff,i}$ can be defined as \cite{Hirschorn2010,Diard2013}:\\

\begin{equation}
\upomega_{c,i}=1/\tau_{i}=(R_{i}Q_{i})^{-1/{\upalpha_{i}}}
\end{equation}

\begin{equation}
C_{eff,i}=Q_{i}^{1/\upalpha_{i}}R_{i}^{1/\upalpha_{i}-1}
\end{equation}

\subsection{Impedance Evolution with Growth and Distance}

\begin{figure}[h]
  \centering
  \includegraphics[width=1\columnwidth]{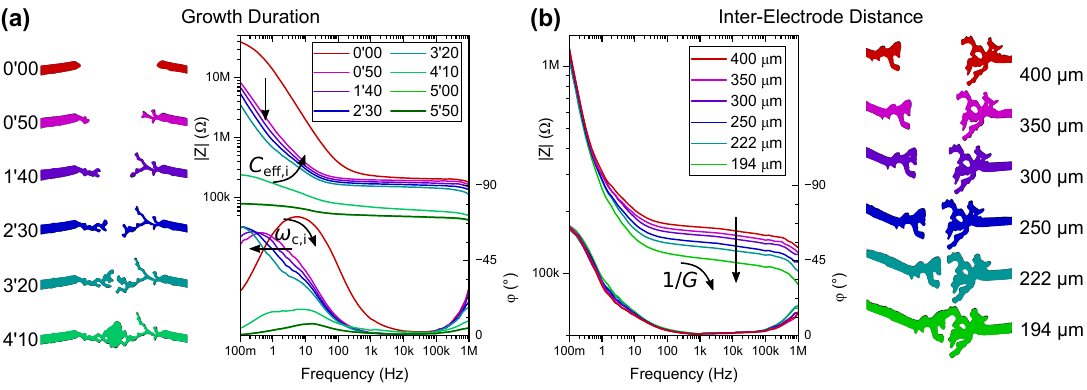}
  \caption{Bode plots in magnitude and phase of various CPD growths, for which impedance measurements were performed intermittently while interrupting a single growth at fixed time intervals (a), and by varying the inter-electrode distance after a single growth with the micro-manuplators holding the gold wires (b). For all cases, binarized/vectorized images of the different morphologies are shown with the label corresponding to the related duration or distance.}
  \label{fig:fig4}
  \end{figure}

EIS performed between two gold wires featuring two disconnected CPD systematically show a high-pass filter for all morphologies (see Fig.~\ref{fig:fig4}). The frequency cutoff lies systematically below 1 kHz, and the decrease in impedance modulus |\textit{Z}| occurs at a much lower rate than one decade of |\textit{Z}| per decade of \textit{f}. In addition, the phase $\upphi$ never reaches -90$^{\circ}$ in the 3-4 decades below the cutoff, despite the fact that no low-frequency plateau for |\textit{Z}| is observed. This clearly shows that a conventional serial RC would not be suitable as a circuit model to fit the experimental impedance data. Furthermore, the relaxation dynamics seems to show non-ideal traces compared to a charging-discharging process of a capacitor described by an exponential function, following a Debye relaxation model. In this study, the distance between two CPD has been changed (Fig.~\ref{fig:fig4}b) in order to quantify the contribution of the Inter-Electrode Distance (\textit{L}) in the decrease of the |\textit{Z}| response when two CPD grow toward each other (Fig.~\ref{fig:fig4}a) decreasing \textit{L} as the morphology is evolving. During growth, we observed that |\textit{Z}| diminishes at all frequencies (Fig.~\ref{fig:fig4}a) while narrowing the distance between two CPD without modifying their structure has only an impact on |\textit{Z}| at high frequencies. This result highlights several properties. First, the systematic change in impedance with the growth duration clearly reduces the impact the mm-sized gold wires have in the system relaxation, when compared to the contribution of the micrometer-scaled CPD. It is even observed that the impedance is substantially changed at the beginning of the experiment when the first tenths of micrograms are deposited on the apex of the wires. Second, narrowing the distance between CPD testifies on the fact that the total volume of a CPD dictates the signal propagation and not just the distance between them, which seems to impact only on the electrolyte conductance \textit{G} (Fig.~ \ref{fig:fig4}b). One can notice that the phase $\upphi$ is completely unaffected by the variation of \textit{L}, which shows that reducing the distance has absolutely no impact on frequency-dependent circuit elements in the impedance model. When both CPD finally close the gap and merge into a single CPD (Fig.~\ref{fig:fig4}a, from 3'20), a significant decrease of the overall impedance is observed, until not further variation is found (traces at 5'00 and 5'50 are almost identical). Despite both CPD merging into an electrically conductive PEDOT:PSS structure, a pseudo-capacitive dependency of the CPD is still observed in |\textit{Z}| and $\upphi$, which shows that even if a conductive path is made between two nodes connected by a CPD, the signal still partially propagates through the electrolyte. While the whole spectrum is affected by the growth, the lower frequency region is much more impacted than the higher one. Therefore, one can conclude that growing CPD affects more their impedance by increasing their volume than by decreasing the distance between them. The volume of PEDOT:PSS conditions the specific capacitance as an OMIEC,\cite{Rivnay2015} which in turns modify the (R|CPE) Voigt elements (or ZARC element) in our impedance model (Fig.~\ref{fig:fig4}a). In phase, one can observe that multiple relaxations take place in the spectra when CPD are present on the wires (Fig.~\ref{fig:fig4}a). This demonstrates that at least two frequency-dependent elements are involved in the signal propagation through the CPD. A rise in phase can also be seen around 1~MHz due to the geometric capacitance in our system (parallel to 1/\textit{G}), which is not accounted in our model to focus specifically on modeling CPD's response at low frequencies.\\\\

\subsection{Circuit Element Evolution with the Growth}

\begin{figure}[!h]
  \centering
  \includegraphics[width=1\columnwidth]{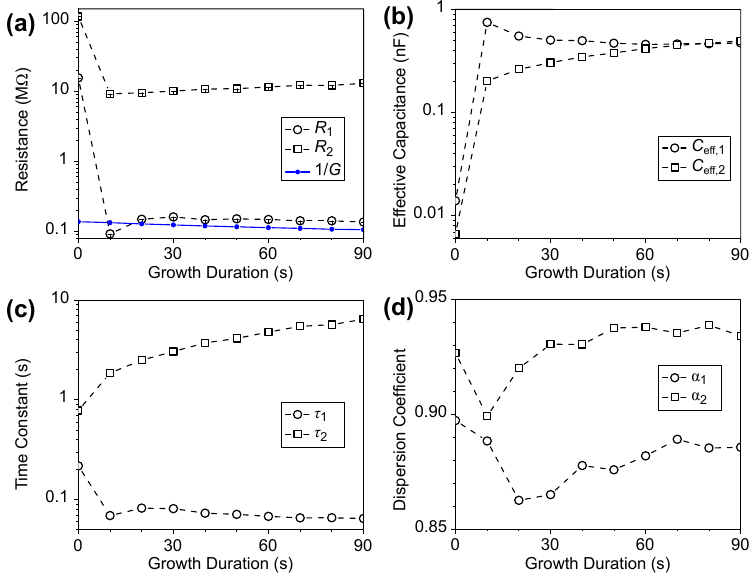}
  \caption{Model's parameters evolution with CPD growth: resistances \textit{R}$_{i}$ or 1/\textit{G} (a), effective capacitances \textit{C}$_{eff,i}$ (b), characteristic time constants $\tau_{i}$ (c) and the dispersion coefficients $\upalpha_{i}$ (d).}
  \label{fig:fig5}
  \end{figure}

A first attempt to fit all impedances using a simplified one-Voigt circuit (similar to the model in Fig.~\ref{fig:fig3}, but accounting for \textit{R}$_{1}$~=~\textit{R}$_{2}$, \textit{Q}$_{1}$~=~\textit{Q}$_{2}$ and $\upalpha_{1}$~=~$\upalpha_{2}$) was very unsatisfying as a model since a single (R|CPE) Voigt circuit representation could not converge to the typical phase features observed in Fig.~\ref{fig:fig4}a (where the left pointing arrow is located). Therefore, the two-Voigt model was used for the fitting without cross-correlating the seven parameters of the model. For different cases, parameters for the naked wire (growth duration: 0~s) were substantially different from the one at different growth stages (Fig.~\ref{fig:fig5}). Charge-transfer resistances \textit{R}$_{i}$ are one to two orders of magnitude higher prior growth and barely evolves with time after the initial decrease (Fig.~\ref{fig:fig5}a). Surprisingly, both \textit{R}$_{1}$ and \textit{R}$_{2}$ differ by about two orders of magnitude. This is the first evidence suggesting that both Voigt subcircuits may not be attributed to each individual CPD, but to different mechanisms common for both CPD. Particularly, \textit{R}$_{1}$ (related to the slower event as $\tau_{1}$~> 0.5~s upon growth) is nearly as resistive as the electrolyte 1/\textit{G} near 100~k$\Upomega$ whereas \textit{R}$_{2}$ (related to the faster event as $\tau_{2}$ < 0.5~s upon growth) is at the level of 10~M$\Upomega$ (Fig.~\ref{fig:fig4}a). 1/\textit{G} decreases quasi linearly from 138~k$\Upomega$ (0~s) to 106~k$\Upomega$ (90~s). The effective capacitances \textit{C}$_{eff,i}$ of both relaxations increase from 10±4~pF for naked wires to 140±70~pF at the first measurement during the growth, to converge both to the same value at 470±40~pF after 70~s (Fig.~\ref{fig:fig5}b). The fact that both relaxations differ by at least a factor of 10$\times$ upon growth (Fig.~\ref{fig:fig5}c) with a significant difference on the charge-transfer resistance (Fig.~\ref{fig:fig5}a) compared to the effective capacitance (Fig.~\ref{fig:fig5}b) may suggest that the limiting mechanisms for signal propagation between adjacent CPD is influenced more by the ionic transport through their morphology rather than the actual volume a CPD occupies. The dispersion coefficients $\upalpha_{i}$ of the CPE$_{i}$ are also monitored over growth (Fig.~\ref{fig:fig5}d) and do not show substantial variations over time. Moreover, all values for $\upalpha_{1}$ (slow relaxation) and $\upalpha_{2}$ (fast relaxation) remain at 0.90±4 during the whole growth suggesting that the non-ideality of both relaxations does not depend on the quantity of material deposited on the electrode, but is rather intrinsic to a particular topology grown in a given environment.\\

\section{Conclusion}

A first study on the impedance monitoring of CPD grown at different stages is presented. Despite the fact that these structures could be compared to electrochemical capacitors, their behaviour is rather more specific. The impedance evolution is not particularly impacted by their distance nor the volume they occupy. In fact, both evolving relaxations are rather depending on the evolution of the conductive component and less the pseudo-capacitive one. This suggests that a CPD structure conditions differently the impedance depending on its topological property, such as fractality, number of branches or their directions, and less on size/distance or volume. As the dispersion coefficient of a CPE may be associated to such topological structures \cite{CordobaTorres2015}, a given CPD morphology may engrave a specific impedance filter to condition the signal transport between two electrodes. This first study encourages the investigation of different topologies (mono-wire, multi-filaments, directional or isotropic dendrites) to verify if these objects are suitable candidates to manufacture topological electronic circuitries with a low amount of ressources, exploiting bio-inspired information processing strategies.\\

\section*{Acknowledgments}
The authors thank the French National Nanofabrication Network RENATECH for financial support of the IEMN cleanroom. We thank also the IEMN cleanroom staff for their advice and support. This work is funded by ANR-JCJC "Sensation" project (grant number: ANR-22-CE24-0001).\\

\bibliographystyle{natsty-doilk-on-jour}  
\small
\bibliography{main}

\end{document}